# Ultrafast all-optical shutter based on two-photon absorption

**Marijn A. M. Versteegh and Jaap I. Dijkhuis***

*Debye Institute for Nanomaterials Science, Utrecht University, Princetonplein 1, 3584 CC Utrecht, The Netherlands*
**Corresponding author: j.i.dijkhuis@uu.nl*

An ultrafast all-optical shutter is presented, based on a simple two-color two-photon absorption technique. For time-resolved luminescence measurements this shutter is an interesting alternative to the optical Kerr gate. The rejection efficiency is 99%, the switching-off and switching-on speeds are limited by the pulse length only, the rejection time is determined by the crystal slab thickness, and the bandwidth spans the entire visible spectrum. We show that our shutter can also be used for accurate measurement of group velocity inside a transparent material.

Development of ultrafast optical gates and shutters is of enormous importance for studies on ultrafast carrier dynamics and relaxation processes, lasing, ultrafast acoustics, charge transport on the nanoscale, light diffusion, and transient excited states occurring in chemical reactions. In addition, ultrafast optical gating has many potential applications in integrated optics and ultrahigh-speed information processing.

An often used method for optical gating is based on sum-frequency generation [1]. In a nonlinear crystal the signal beam is mixed with a femtosecond gating pulse. The time resolution is in principle limited by the pulse length only, but the efficiency is limited, and this method only works for one wavelength at a time, since the phase-matching condition must be satisfied.

Another powerful method is based on the optical Kerr effect [2]. A femtosecond pulse induces birefringence in a medium. The signal beam traversing this Kerr medium experiences a polarization rotation. With the use of polarizers this effect serves to either block the beam during a short time interval, creating the optical Kerr shutter, or to let it pass, creating the optical Kerr gate. Optical Kerr gating has the advantage that the phase-matching condition is automatically fulfilled for each wavelength, and that the efficiency can be increased to 5-10% [3,4]. On the other hand, the time resolution generally is not just limited by the pulse length, but also by the response time and the relaxation time of the Kerr medium. For $CS_2$, for example, the response time is 0.8 ps. Since the advent of femtosecond lasers several materials have been found or made with a faster Kerr response [3-7]. Also novel materials such as photonic crystals are employed as ultrafast switch [8,9].

In this Letter, we present an ultrafast all-optical shutter based on two-photon absorption in ZnO. Our measurements show a rejection efficiency of 99%. The switching-on and -off time resolution is limited by the pulse length only. The rejection time is determined by the crystal thickness. Phase matching is not required and the technique operates over the full visible spectrum, between 385 nm and 700 nm.

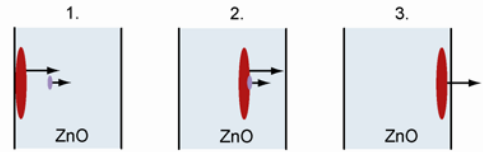

Fig. 1. (Color online) Principle of the ultrafast optical shutter. The small pulse represents the signal pulse; the large pulse is the gating pulse.

The principle of our shutter is explained in Fig. 1. The light pulse to be measured, the signal pulse, is sent through a ZnO single crystal slab. It is followed by an 800-nm gating pulse with a higher travel speed and intensity than the signal pulse. The gating pulse catches up with the signal pulse, leading to two-color two-photon absorption and annihilation of the signal pulse.

This shutter method uses the fact that ZnO has a large direct band gap of 3.37 eV at room temperature, making the semiconductor transparent for visible light. When the 800-nm gating pulse meets the signal pulse, two-photon absorption occurs for signal wavelengths below 700 nm. Since the group index of refraction of ZnO at 800 nm is lower than at visible wavelengths, the gating pulse always propagates the fastest.

Two-photon absorption of course only occurs if the delay between the two pulses is such that they overlap inside the crystal. If the signal pulse arrives before the gating pulse, or if it has left the crystal before being caught up by the gating pulse, it transverses the ZnO slab unhindered. The width of the delay domain for which two-photon absorption takes place, is determined by the speeds of the two pulses and by the thickness of the crystal.

For our experiment, 140-fs (full-width half-maximum) 800-nm laser pulses from an amplified 1-kHz Ti:sapphire laser were split into a weak signal pulse and a strong gating pulse. Employing self-phase modulation in a sapphire crystal, the 800-nm signal pulses were converted into white-light pulses, enabling measurements in the visible down to 450 nm. The signal pulses were sent through a color filter to block 800-nm light and focused into an epi-polished 523-μm thick ZnO single crystal oriented in the [0001] direction, i.e. with the *c*-axis perpendicular to the wafer plane. The 800-nm gating pulses were sent through a delay line and focused in such a way that the gating track was fully overlapping the signal track through the crystal. For both signal pulse and gating pulse the polarization was $\mathbf{E} \perp \mathbf{c}$. A grating and a slit were used to select the wavelength to be measured, with a bandwidth of 4 nm. Finally the transmitted signal light was measured by a photodiode as a function of delay with respect to the gating pulse.

We have also performed measurements at signal wavelengths around 400 nm. For these measurements a BBO crystal was placed behind the sapphire. There, by sum-frequency generation of 800-nm light with one of the frequencies present in the white-light pulse a signal pulse was created with a wavelength between 390 and 425 nm and a bandwidth of 2 nm (FWHM). By changing the orientation of the BBO crystal the wavelength could be tuned. Again, color filters were used to block unwanted wavelengths before the pulse was sent through the ZnO crystal.

All measurements were performed with a signal fluence much smaller than the gating fluence. Under this condition the experimental data do not depend on the signal fluence.

value. This result implies that also for negative delays reflected gating light must have a minor contribution to the rejection of the signal light.

In our experiment we used short signal pulses. The switching-off speed depends on the gating and signal pulse lengths at the back side of the ZnO crystal, the switching-on speed on those at the front side. Because of dispersion inside the ZnO, switching-off therefore is slower than switching-on. This behavior is clearly observed in Fig. 2, especially for the shortest wavelengths, where the dispersion is the strongest.

If one wishes to measure a longer signal pulse, a 10-ps luminescence signal for instance, then this ultrafast shutter is very suitable to measure it in a time-resolved way. Suppose that the luminescence wavelength is 425 nm. The shutter then annihilates the luminescence for a period of 2.8 ps. By varying the delay in small steps one can with high resolution determine the time-dependent luminescence intensity.

The width $w$ of the shutter boxcar (the rejection time) depends on the wavelength-dependent group velocities of signal light and gating pulse, and the thickness of the crystal $L$ as

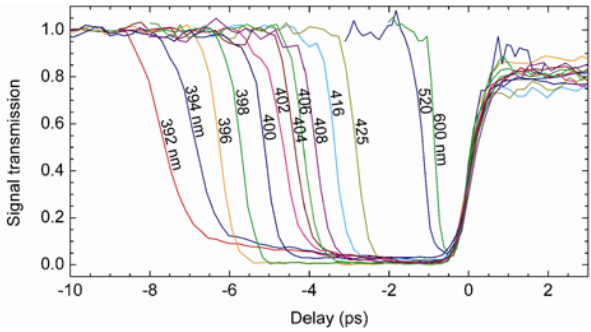

Fig. 2. (Color online) Ultrafast optical shutter. Measured transmitted signal light versus delay with respect to the gating pulse for indicated signal wavelengths. Transmission has been normalized to the measured value without gating pulse. Gating fluence was 210 J/m², gating pulse duration 140 fs, and slab thickness 523 μm.

The results, shown in Fig. 2, are boxcar signals with sharp edges, proving the ultrafast shutter effect. For example, 425-nm signal light arriving less than 2.8 ps before the gating pulse is caught up and annihilated by the gating pulse via two-photon absorption. Light arriving more than 2.8 ps before the gating pulse is not overtaken inside the ZnO crystal and therefore unaffected. The same applies for signal light arriving after the gating pulse. The transmission inside the dip is around or below 1%, showing 99% rejection efficiency of the shutter.

We observe residual two-photon absorption for positive delays: at the right side of the boxcars in Fig. 2 the transmission is at a lower level than at the left side. This is explained by 10.5% of the gating pulse intensity which is reflected at the back side of the crystal. For large positive delays the transmission returns to its initial

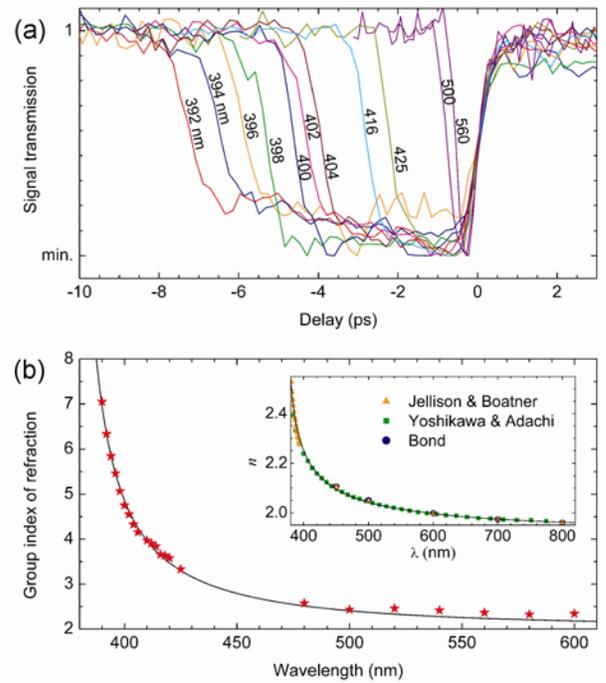

Fig. 3. (Color online) Measurement of the group index of refraction in ZnO using the ultrafast optical shutter. (a) Signal transmission versus delay with respect to the gating pulse for indicated signal wavelengths at a gating fluence of 10 J/m². (b) Red stars: group refractive index derived from the measured boxcar widths. Line: group refractive index calculated from the phase refractive index shown in the inset. Data points in the inset are from Refs. 10-12.

$$w(\lambda) = \frac{L}{v_g(\lambda_{\text{signal}})} - \frac{L}{v_g(\lambda_{\text{gate}})}. \qquad (1)$$

The group velocity $v_g(\lambda)$ here equals $c/n_g(\lambda)$, where $c$ is the speed of light in vacuum, and $n_g(\lambda)$ is the group index of refraction, related to the phase index of refraction $n(\lambda)$ as [13]

$$n_g(\lambda) = n(\lambda) - \lambda \frac{dn(\lambda)}{d\lambda}. \qquad (2)$$

In ZnO, both $n(\lambda)$ and $n_g(\lambda)$ are monotonously decreasing functions in the wavelength range considered here. Therefore the 800-nm gating pulse traverses the crystal faster than a signal pulse in the visible does.

When the group refractive index at the gating wavelength is known, the ultrafast optical shutter can be used to measure the group refractive index for a selected signal wavelength. One should however be vigilant for saturation effects: the data at high gating pulse fluences shown in Fig. 2 exhibit a slight broadening of the dip beyond the values following from Eq. (1). For accurate measurement of the group refractive index via the rejection time, we therefore performed an experiment at a lower gating fluence, where the rejection of the signal pulse is around 50%. Results are shown in Fig. 3(a). Indeed, compared to Fig. 2 the dips are slightly narrower.

In order to prove that this method yields faithful results, we compare the measurements in Fig. 3(b) with experimental data in literature on the phase index of refraction of ZnO at room temperature. From the line drawn in the inset, the group index $n_g(\lambda)$ is calculated using Eq. (2), and the result is the line in Fig. 3(b). For the gating pulse we find $n_g(800 \text{ nm}) = 2.04$. On the other hand, the measured rejection times, with the use of Eq. (1), yield the red stars. Here we inserted the value of 2.04 for the gating group index. The agreement is remarkably good, proving that the ultrafast optical gate is able to yield accurate results for the group index of refraction. Only for signal wavelengths longer than 500 nm, the data are slightly above the literature line. This can be understood from the fact that for these wavelengths the velocity difference between gate and signal is small and the left and right sides of the dip are not well separated anymore, so that Eq. (1) becomes less applicable. In such cases a thicker sample would yield better results.

To conclude, we have presented an ultrafast all-optical shutter based on two-color two-photon absorption in a ZnO crystal and accurately measured its group index dispersion. The shutter very rapidly switches off a light signal and after a few picoseconds very quickly switches it on again. The rejection time of the ultrafast switch can be controlled by the thickness of the crystal slab. Our measurements show a 99% reduction of the light transmission by the shutter. The switching-off and switching-on speeds are limited by the pulse length only.

We believe that our optical shutter is promising for applications since it is broad band and phase matching is not needed. In combination with a delay line and lock-in techniques this shutter should be very suitable for measurement of time-resolved intensity profiles of weak light pulses. Other applications one could think of are suppression of spurious Rayleigh laser light and reduction of detection noise in ultrafast pump-probe experiments, or pulse shaping in coherent optics.

We thank T. Kuis for help with the measurements and we thank C. R. de Kok and P. Jurrius for technical support.

### References

1. D. Block and J. Shah, "Femtosecond luminescence measurements in GaAs," Solid State Commun. **59,** 527-531 (1986).
2. M. A. Duguay and J. W. Hansen, "An ultrafast light gate," Appl. Phys. Lett. **15,** 192-194 (1969).
3. J. Takeda, K. Nakajima, S. Kurita, S. Tomimoto, S. Saito, and T. Suemoto, "Time-resolved luminescence spectroscopy by the optical Kerr-gate method applicable to ultrafast relaxation processes," Phys. Rev. B **62,** 10083-10087 (2000).
4. B. L. Yu, A. B. Bykov, T. Qiu, P. P. Ho, R. R. Alfano, and N. Borrelli, "Femtosecond optical Kerr-shutter using lead-bismuth-gallium oxide glass," Opt. Commun. **215,** 407-411 (2003).
5. D. Hulin, J. Etchepare, A. Antonetti, L. L. Chase, G. Grillon, A. Migus, and A. Mysyrowicz, "Subpicosecond time-resolved luminescence spectroscopy of highly excited CuCl," Appl. Phys. Lett. **45,** 993-995 (1984).
6. S. Kinoshita, H. Ozawa, Y. Kanematsu, I. Tanaka, N. Sugimoto, and S. Fujiwara, "Efficient optical Kerr shutter for femtosecond time-resolved luminescence spectroscopy," Rev. Sci. Instrum. **71,** 3317-3322 (2000).
7. S. Tatsuura, T. Matsubara, H. Mitsu, Y. Sato, I. Iwasa, M. Tian, and M. Furuki, "Cadmium telluride bulk crystal as an ultrafast nonlinear optical switch," Appl. Phys. Lett. **87,** 251110 (2005).
8. Y. Liu, F. Qin, Z. Y. Wei, Q. B. Meng, D. Z. Zhang, and Z. Y. Li, "10 fs ultrafast all-optical switching in polystyrene nonlinear photonic crystals," Appl. Phys. Lett. **95,** 131116 (2009).
9. D. A. Mazurenko, R. Kerst, J. I. Dijkhuis, A. V. Akimov, V. G. Golubev, D. A. Kurdyokov, A. B. Pevtsov, and A. V. Sel'kin, "Ultrafast Optical Switching in Three-Dimensional Photonic Crystals," Phys. Rev. Lett. **91,** 213903 (2003).
10. G. E. Jellison, Jr. and L. A. Boatner, "Optical functions of uniaxial ZnO determined by generalized ellipsometry," Phys. Rev. B **58,** 3586-3589 (1998).
11. H. Yoshikawa and S. Adachi, "Optical constants of ZnO," Jpn. J. Appl. Phys. **36,** 6237-6243 (1997).
12. W. L. Bond, "Measurement of the Refractive Indices of Several Crystals," J. Appl. Phys. **36,** 1674-1677 (1965).
13. J. D. Jackson, *Classical Electrodynamics,* 3rd. ed. (Wiley, New York, 1998).